\documentstyle[euscript,12pt]{article}
\catcode`@=11
\newcount\@tempcntc
\def\@citex[#1]#2{\if@filesw\immediate\write\@auxout{\string\citation{#2}}\fi
  \@tempcnta\z@\@tempcntb\m@ne\def\@citea{}\@cite{\@for\@citeb:=#2\do
    {\@ifundefined
       {b@\@citeb}{\@citeo\@tempcntb\m@ne\@citea\def\@citea{,}{\bf ?}\@warning
       {Citation `\@citeb' on page \thepage \space undefined}}%
    {\setbox\z@\hbox{\global\@tempcntc0\csname b@\@citeb\endcsname\relax}%
     \ifnum\@tempcntc=\z@ \@citeo\@tempcntb\m@ne
       \@citea\def\@citea{,}\hbox{\csname b@\@citeb\endcsname}%
     \else
      \advance\@tempcntb\@ne
      \ifnum\@tempcntb=\@tempcntc
      \else\advance\@tempcntb\m@ne\@citeo
      \@tempcnta\@tempcntc\@tempcntb\@tempcntc\fi\fi}}\@citeo}{#1}}
\def\@citeo{\ifnum\@tempcnta>\@tempcntb\else\@citea\def\@citea{,}%
  \ifnum\@tempcnta=\@tempcntb\the\@tempcnta\else
   {\advance\@tempcnta\@ne\ifnum\@tempcnta=\@tempcntb \else \def\@citea{--}\fi
    \advance\@tempcnta\m@ne\the\@tempcnta\@citea\the\@tempcntb}\fi\fi}
\catcode`@=12
\def\be{\begin{equation}}
\def\ee{\end{equation}}
\def\barr{\begin{array}}
\def\earr{\end{array}}
\def\bea{\begin{eqnarray}}
\def\eea{\end{eqnarray}}
\def\bmath{\begin{displaymath}}
\def\emath{\end{displaymath}}
\def\bq{\begin{quote}}
\def\eq{\end{quote}}
\def\oas{$\EuScript O(\alpha_s)$}
\def\Tr{\mbox{$\mbox{\rm Tr}\;$}}
\def\g5{\gamma_5}
\def\as{\alpha_s}
\def\real{\mathop{\mbox{\rm Re}}\nolimits}
\def\imag{\mathop{\mbox{\rm Im}}\nolimits}
\def\gE{\gamma_{\scriptscriptstyle E}}
\def\chiz{\chi_{\scriptscriptstyle Z}}
\def\mz{M_{\scriptscriptstyle Z}}
\def\gz{\Gamma_{\scriptscriptstyle Z}}
\def\gf{g_{\scriptscriptstyle F}}
\def\CF{C_{\scriptscriptstyle F}}
\def\NC{N_{\scriptscriptstyle C}}
\def\ct{\cos\theta}
\def\c2t{\cos^2\kern-2pt\theta}
\def\st{\sin\theta}
\def\s2t{\sin^2\kern-2pt\theta}
\def\su{\sigma_{\scriptscriptstyle U}}
\def\sl{\sigma_{\scriptscriptstyle L}}
\def\sf{\sigma_{\scriptscriptstyle F}}
\def\Li{\mbox{$\mbox{\rm Li}_2$}}
\def\Frac#1#2{\mbox{$\textstyle{#1\over#2}$}}
\def\eps{\varepsilon}
\def\nn{\nonumber\\}

\def\Slash#1{\mbox{$\not{\hspace{-1.03mm}#1}$}}
\def\bm#1{\mbox{$\boldmath{#1}$}}

\def\direc{\bm{\hat{p}}}
\def\ndirec{\bm{\hat{n}}}
\def\I#1{{\EuScript I}_{#1}}
\def\S#1{{\EuScript S}_{#1}}
\def\J#1{{\EuScript J}_{#1}}
\def\T#1{{\EuScript T}_{#1}}
\def\II#1{{\tilde{\EuScript I}}_{#1}}
\def\SS#1{{\tilde{\EuScript S}}_{#1}}
\def\JJ#1{{\tilde{\EuScript J}}_{#1}}
\def\uint{\displaystyle\int dy\,dz\:\:}
\def\pint{\displaystyle\int{dy\,dz\over\sqrt{(1-y)^2-\xi}}\:\:}
\def\jint{\displaystyle\int{dy\,dz \over (1-y)^2-\xi}\:\:}

\def\sxi{\mbox{$\sqrt\xi$}}

\begin{document}
\thispagestyle{empty}
\begin{flushright}
FTUV/95-38 \\[-0.2cm]
IFIC/95-40 \\[-0.2cm]
February 1995 \\[-0.2cm]
(revised version) \\[-0.2cm]
\end{flushright}
\begin{center}

{\bf{\Large Analytic {\boldmath\oas} Results for}}\\[.3cm]
{\bf{\Large Bottom and Top Quark Production}}\\[.3cm]
{\bf{\Large {in \boldmath$e^+e^-$} Collisions}}\\[1.25cm]

{\large
M.M.~Tung~\footnote{Feodor-Lynen Fellow},
J.~Bernab\'eu, and J.~Pe\~narrocha
} \\[.4cm]
Departament de F\'\i sica Te\`orica, Universitat~de Val\`encia \\
and IFIC, Centre Mixte Universitat Val\`encia --- CSIC, \\
C/ Dr.~Moliner, 50, E-46100 Burjassot (Val\`encia), Spain.
\end{center}
\vspace{3cm}
\centerline {\bf ABSTRACT}
\noindent
We present a new derivation of the \oas\ angular distribution of the
outgoing $q$-quark in the production process $e^+ e^- \to\gamma,Z\to
q\,\bar{q}(g)$. In our calculation, we express the three-particle phase-space
integration of the gluon-bremsstrahlung process in terms of a general set of
analytic integral solutions. A consistent treatment of the QCD one-loop
corrections to the axial-vector current deserves special attention. This is
relevant in the derivation of the forward-backward asymmetry predicted by the
standard model. Finally, we provide the full analytical solutions for the
differential rates in closed form and conclude with numerical estimates for
bottom and top quark production.
%\\[1cm]
%\noindent PACS number(s): 11.15.Bt, 12.38.Bx, 13.38.+c \\

\newpage
\noindent

\section{Introduction}
The recent discovery of the top quark has intensified interest for radiative
corrections in the Standard Model. In particular, processes involving
top pair production are very attractive for closer investigation due
to the absence of fragmentation effects. For bottom quark production,
the measurement of the forward-backward asymmetry from $Z$-decays allows
for a very precise determination of the electroweak mixing angle and
the vector/axial-vector couplings to the fermions. The high accuracy of
these experiments as carried out by the LEP~\cite{LEP} and SLC~\cite{SLC}
collaborations requires theoretical predictions beyond Born level in
the perturbative expansion of electromagnetic and strong couplings.

In this regard, the calculation of differential cross sections for the
electroweak production of massive quarks from $e^+ e^-$ annihilation
is of crucial importance. Radiative corrections to this process are
in general dominated by QCD effects, which can be calculated perturbatively
at energies sufficiently above the production threshold of the quark pair.

A first analytical treatment of radiative corrections to the angular
distributions in heavy-quark production was carried out in Ref.~\cite{grun}
at energies where the $e^+ e^-$ annihilation process is dominated by pure
photon exchange. At higher energies, the $\gamma-Z$ interference has to be
included~\cite{JLZ}. This leads to a considerable complication in the
three-body phase-space (gluon bremsstrahlung off a massive quark pair).
Only recently~\cite{ABL}, expressions for the bremsstrahlung process have
been obtained by including cuts on the energy of the emmitted gluon.

In this article, we present a systematic derivation of the QCD one-loop
corrections to the differential production rate for the annihilation
process $e^+ e^- \to\gamma,Z\to q\,\bar{q}(g)$. No approximations are
made and the whole three-body phase-space is considered. The full dependence
on the quark masses is kept so that the final analytical expressions apply to
the full energy spectrum where perturbative QCD is valid.

Our approach relies on a general set of integral solutions completely
describing the complicated phase-space with two massive fermions and one
vector particle. The method involves special integration techniques and
was devised in Ref.~\cite{thesis}. It was used
in various calculations at \oas\ such as the total longitudinal or
beam-alignment polarization for quarks produced in $e^+ e^-$
annihilation~\cite{long,approx,beam}. In the present work, we have been lead
to an additional different class of phase-space integrals due to the
more intricate angular structure of the differential calculation.
Furthermore, we have used two different $\g5$-prescriptions in the derivation
of the $\EuScript C$-odd part of the differential cross section. There, the
corresponding traces contain an odd number of the Dirac $\g5$ and one faces
immediately the problem on how to treat it consistently within dimensional
regularization.

The article is organized as follows. In Section~2, we set up the basic
formalism for the calculation of the differential cross section
$d\sigma(e^+ e^-\to\gamma,Z\to q\bar{q})/d\ct$ (the angle $\theta$ is the
usual scattering of the $q$-quark with respect to the electron beam axis)
including Born approximation and QCD virtual corrections. Special
attention is paid to the gluonic vertex corrections within different
regularization schemes. In Section~3 follows a detailed presentation of the
kinematics of the massive $q\bar{q}g$ phase-space which directly gives the
angular distributions for the \oas\ tree-graph contributions. Compact
expressions for the real-gluon corrections are derived. The $\EuScript C$-odd
partial rate is calculated in two independent approaches
yielding identical results. In Section~4, we present the full analytical results
for the differential rates and conclude with an explicit numerical analysis for
bottom and top quark production. The article is supplemented by two Appendices
A and B where details on the renormalization of the axial-vector current and
the massive three-body phase-space can be found.
\section{Basic Framework and Virtual Corrections}
The differential cross section for the production process $e^+ e^-\to q\bar{q}$
is a binomial in $\cos\theta$, $\theta$ being the scattering angle of
the tagged quark. It is common to introduce the structure functions
$\su, \sl$, and $\sf$ as follows
\be\label{1}
{d\:\sigma\kern10pt\over d\ct} =
\Frac{3}{8}(1+\c2t)\,\su+\Frac{3}{4}\s2t\,\sl+\Frac{3}{4}\ct\,\sf,
\ee
so that $\su$ and $\sl$ are the contributions stemming from unpolarized
and longitudinally polarized gauge bosons, respectively. The structure
function $\sf$ relates to the difference of left- and right-chiral
polarizations of the quarks and constitutes the $\EuScript C$-odd component
under the exchange of quark and antiquark in the final state. All structure
functions $\sigma_i\ (i=U,L,F)$ contain the electroweak couplings of the
process under consideration.

For the calculation of the different terms in Eq.~(\ref{1}) (corresponding
to Born and virtual corrections as shown in Fig.~1a,b) it is more convenient
to rewrite the differential cross section in terms of all possible
parity-parity combinations $i,j=V,A$ that occur through the virtual
$\gamma$ and $Z$ states:
\bea
&&{d\:\sigma\kern10pt\over d\ct} \ =\
{\NC\over 16}\,\left({\alpha\over q^2}\right)^2 v\,\sum_{i,j=V,A}
g^{ij}\,\int d\varphi\:L^{\mu\nu}\,H_{\mu\nu}^{ij} \nn
&&=\ {3\over 8}\,\pi\,\left({\alpha\over q^2}\right)^2 v\,
\Bigg[\,g^{VV}\,L^{(\mu\nu)}\,H_{(\mu\nu)}^{VV}+g^{V\!A}\,L^{[\mu\nu]}
\,H_{[\mu\nu]}^{V\!A+AV}+g^{AA}\,L^{(\mu\nu)}\,H_{(\mu\nu)}^{AA}\,\Bigg],
\label{master}
\eea
where $\theta$ and $\varphi$ are the the polar and azimuthal angles of the
scattered quark. Here, $L$ and $H$ denote the lepton and hadron tensor,
respectively, and as usual $\NC=3$ takes into account that the produced quark
pair comes in three colors. Because of cylindrical symmetry around the beam
axis (as indicated in Fig.~2) the integration over $\varphi$ is simplest.
The mass parameter $v=\sqrt{1-4m^2/q^2}$ ($m$ quark mass, $q$ total
energy-momentum transfer) enters Eq.~(\ref{master}) via the two-body
phase-space factor $\mbox{PS}_2=v/8\pi$. Note that the product of lepton
and hadron tensor with fixed parities $L^{\mu\nu}\,H_{\mu\nu}^{ij}$ is
multiplied by the corresponding coefficients $g^{ij}$
\bea
g^{VV}   &=&
Q_q^2-2\,Q_q v_e v_q \real{\chiz}+(v_e^2+a_e^2)\,v_q^2\,|\chiz|^2, \\
g^{V\!A} &=& -Q_q a_e a_q \real{\chiz}+2\,v_e a_e v_q a_q |\chiz|^2, \\
g^{AA}   &=& (v_e^2+a_e^2)\,a_q^2\,|\chiz|^2,
\eea
which contain the neutral-current couplings
$v_f=2\,T_z^f-4\,Q_f \sin^2\theta_{\scriptscriptstyle W}$ and $a_f=2\,T_z^f$,
and $Q_f$ denotes the fractional charge of the fermion. The Breit-Wigner
form of the massive gauge boson is characterized by
\be
\chiz(q^2)={\gf\,\mz^2\,q^2\over q^2-\mz^2+i\mz\,\gz}\quad\mbox{and}\quad
\gf={G_{\scriptscriptstyle F}\over 8\sqrt{2}\,\pi\alpha}\approx
4.299\cdot 10^{-5}\,\mbox{GeV}^{-2}.
\ee

The normalization of the symmetric and antisymmetric components in the
tensor decomposition
\be
L^{\mu\nu} = L^{(\mu\nu)} + L^{[\mu\nu]}
\ee
has to be in agreement with Eq.~(\ref{master})
\bea
L^{(\mu\nu)} &=& {4\over q^2}\,\Big(\, p_+^\mu p_-^\nu+p_+^\nu p_-^\mu
                 -\Frac{1}{2}q^2 g^{\mu\nu}\,\Big), \\
L^{[\mu\nu]} &=& {4i\over q^2}\,\epsilon(\mu,\nu,p_-,p_+),
\eea
where $p_-$ and $p_+$ refer to the momenta of the electron and positron,
respectively, and $q=p_- +p_+$ is the total momentum transfer as shown
in Fig.~1b.

In Fig.~2, we show the kinematic configuration of the quark-antiquark
final state in the center-of-momentum system (cms). The initial $e^+ e^-$
beam is aligned along the $z$-axis, so that the polar angle $\theta$
coincides with the scattering angle between the outgoing quark momentum
{\boldmath $p_1$} and the incoming electron momentum {\boldmath $p_-$}.
As usual, the square of the cms energy is given by
$q^2=(p_1+p_2)^2=(p_-+p_+)^2$. Recall that the remaining degree of freedom
is captured by the rotation angle $\varphi$ around the beam axis.
In this coordinate frame, the particle momenta take a particularly
simple contravariant form and the orientation of the quark momentum is given
by $\direc = \Big(\st\,\cos\varphi,\,\st\,\sin\varphi,\,\ct\,\Big)$.
Notice the meaning of $v=\sqrt{1-4m^2/q^2}$ as a dimensionless scale
of the quark's velocity in the cms frame.

In the on-shell renormalization scheme the self-energy corrections vanish so
that only the Feynman diagrams Figs.~1a and 1b are required to calculate the
hadron tensor for the complete \oas\ process with two quarks $q(p_1)$ and
$\bar{q}(p_2)$ in the final state. We choose the Feynman gauge for the internal
gluon propagator. Then, the general substitutions made to arrive
at the massive QCD vertices from the Born term expressions are simply
\bea
\gamma_\mu &\to& (1+A)\,\gamma_\mu-B\,{(p_1-p_2)_\mu\over2m}, \\
\gamma_\mu\g5 &\to& (1+C)\,\gamma_\mu\g5
+D\,{(p_1+p_2)_\mu\over2m}\,\g5,
\eea
with the induction of charge and magnetic moment form factors in the vector
vertex and an axial-charge form factor in the axial vertex.

The explicit real parts of the chromomagnetic form factors are given by
\bea
\real{A} &=& {\as\over 4\pi} \CF\Bigg[\,\Bigg( {1+v^2\over v}
   \ln\Bigg({1-v\over 1+v}\Bigg)+2\Bigg)\Bigg(-\ln\Lambda
   +\ln(1-v^2)-2\ln2\Bigg) \nn && \kern40pt
   +F(v)\,\Bigg], \label{ff1} \\
\real{B} &=& {\as\over4\pi} \CF
   {1-v^2\over v}\ln\Bigg({1-v\over1+v}\Bigg), \label{ff2} \\
\real{C} &=& \real{A}-2\,\real{B}, \label{ABC}
\eea
where $\Lambda$ is a cut-off parameter to control the soft infrared
divergences. At one-loop order, the cutoff $\Lambda$ is related to the
usual dimensional regulator $\eps=4-N$ by the following correspondence
rule~\cite{muta}
\be
\ln\Lambda\,\longleftrightarrow\,{2\over4-N}-
\gE+\ln\left({4\pi\mu^2\over q^2}\right),
\ee
where $\gE$ is the Euler constant and $\mu$ the 't Hooft mass associated
to the couplings in $N$ dimensions. Furthermore, the shorthand $F(v)$ was
introduced
\be\label{F}
F(v) = {1+v^2\over v}\Bigg[\,\Li\Bigg({2v\over v-1}\Bigg)-
       \Li\Bigg({2v\over v+1}\Bigg)+\pi^2\,\Bigg]
       +3\,v \ln\Bigg({1+v\over 1-v}\Bigg)-4.
\ee
As usual $\CF=4/3$ is the Casimir operator defining the adjoint representation
of the $\mbox{SU}(3)_c$ color group. Note that the identity $C = A - 2\,B$
connects the gluonic vertex corrections of the vector and axial-vector
currents for massive fermions.

The form factors Eqs.~(\ref{ff1})--(\ref{ABC}) have been derived in
Ref.~\cite{JLZ} with a completely anticommuting Dirac $\g5$ in dimensional
regularization following the prescription of Chanowitz
{\it et al.\/}~\cite{CFH}. The same final expressions~\cite{long,red} are
also obtained by employing dimensional reduction. In this particular method,
the Clifford algebra of the Dirac matrices is reduced to 4 dimensions while
space-time is still kept in $N=4-\eps$ dimensions to regularize the otherwise
divergent loop integrals. At first glance, naive dimensional reduction and
dimensional regularization seem to differ by a finite term. However, at one-loop
order the inclusion of a global counterterm effectively restores the reduction
in the spin degrees of freedom, so that both schemes are consistent with each
other~\cite{red}.

In dimensional reduction, a straightforward calculation now gives the
\oas\ virtual corrections to the Born approximation. The key ingredients
of Eq.~(\ref{master}) are
\bea
L^{(\mu\nu)}\,H_{(\mu\nu)}^{VV}(virtual\kern1pt) &=&
4\,q^2\,\left[\,(1+2\real{A})\left\{2-v^2(1-\c2t)\right\}\right. \nn
&&\kern1cm\left.+2\real{B}\,v^2(1-\c2t)\,\right], \label{r1} \\
L^{[\mu\nu]}\,H_{[\mu\nu]}^{V\!A+AV}(virtual\kern1pt) &=&
16\,q^2\,(1+\real{A}+\real{C})\,v\,\ct, \label{red1} \label{r2} \\
L^{(\mu\nu)}\,H_{(\mu\nu)}^{AA}(virtual\kern1pt) &=&
4\,q^2\,(1+2\real{C})\,v^2(1+\c2t). \label{r3}
\eea

The soft IR divergences emerge in the QCD form factors $A$ and $C$ as
the cut-off parameter $\Lambda\to 0$, whereas $B$ gives only a finite
correction. The imaginary part $i\imag{D}$ in $H^{AA}_{(\mu\nu)}$
can not contribute to the cross section and finally disappears because of
leptonic current conservation. Note that the contractions of the symmetric
lepton and hadron tensors Eqs.~(\ref{r1}) and (\ref{r3}) contribute to the
$\EuScript C$-even structure functions $\sigma_{\scriptscriptstyle L,U}$,
whereas the contraction of the asymmetric tensors Eq.~(\ref{r2}) yields
$\sf$ which is $\EuScript C$-odd.

At this point, we stress that in the explicit \oas\ calculation of the
$\EuScript C$-odd contribution $\sf$, one faces immediately the well-known
problem of how to extend the four-dimensional $\g5$-matrix to $N$ dimensions,
or, in other words, how to treat the Lorentz indices of the totally
skew-symmetric tensor $\epsilon^{\mu\nu\rho\sigma}$ in $N$ dimensions.
Neither does dimensional regularization nor dimensional reduction
avoid a direct confrontation with the $\g5$-problem: in both cases space-time
is $N$-dimensional. Moreover, the global reduction counterterm of
Ref.~\cite{red} had been taken directly from the massless vector-current vertex.
While its unrestricted usage for loops with even $\g5$ has been demonstrated,
there still remains to investigate its significance for odd $\g5$ calculations.

On the other side, there exists a well-established technique based on the
following substitution of the axial-vector current~\cite{thooft,AD}
\be\label{rep}
\gamma_\mu\g5\,\to\,Z_5\,{i\over 3!}\,
\epsilon_\mu^{\hphantom{\mu}\rho\sigma\tau}\,
\gamma_\rho\gamma_\sigma\gamma_\tau
\ee
with
\be\label{Z5}
Z_5 = 1-{\as\over\pi}\,\CF+\EuScript O(\as^2).
\ee
The finite renormalization $Z_5$ is needed in addition to the conventional
wavefunction renormalization $Z_2$ to restore anticommutativity~\cite{larin}
which is evidently violated after making the replacement Eq.~(\ref{rep}).
Although this replacement method has been developed in first place for massless
multiloop calculations (see e.g.\ \cite{LV}), the generalization to the massive
case is straightforward, since the renormalization of the axial current is
related with the ultraviolet sector, whereas the collinear and soft divergences
relate to the distinct infrared sector.

To complement our calculation, we examine the effects that this particular
regularization scheme has on the gluonic correction of the axial-vector
current. For this purpose, we consider the contraction of the epsilon tensor
with the chiral vacuum polarization tensor (with cuts in the two final fermion
lines of the corresponding diagram). This gives the following definition of
the $C$ form factor:
\bea\label{C1}
C &=& {2\pi\over(q^2)^2 v^2}\,\as\CF\,\epsilon(\mu,\nu,p_1,p_2) \\
  & &
  \times\kern5pt\int{d^N k\over(2\pi)^N}
  {\Tr (\Slash{p_1}+m)\,\gamma_\alpha\,(\Slash{p_1}+\Slash{k}+m)\,
  \gamma_\mu\g5\,(-\Slash{p_2}+\Slash{k}+m)\,\gamma^\alpha\,
  (\Slash{p_2}-m)\,\gamma_\nu \over
  k^2\,\Big\{(p_1+k)^2-m^2\Big\}\,\Big\{(p_2-k)^2-m^2\Big\} },
  \nonumber
\eea
where $k$ is the running momentum of the gluon loop.

Then, we can show that within the framework of the $\g5$-replacement
prescription Eq.~(\ref{rep}) the previous $C$ form factor Eq.~(\ref{ABC})
is recovered
\bea\label{C5}
\real{C} &=&
{\as\over 4\pi} \CF\Bigg[\,2\Bigg( {1+v^2\over v}
   \ln\Bigg({1+v\over 1-v}\Bigg)-2\Bigg)\Bigg(\ln\Lambda^{1\over2}
   -\Frac{1}{2}\ln(1-v^2)+\ln2+1\Bigg) \nn && \kern40pt
   -4v\ln\Bigg({1+v\over1-v}\Bigg)+F(v)+4\,\Bigg].
\eea
In Appendix~A, we outline the derivation of $C$ in Eq.~(\ref{C5}) with a
special emphasis on the renormalization procedure and explicitly point out
the differences to the dimensional reduction approach.

To obtain the result (\ref{C5}) we have used the $\EuScript O(\as)$ expression
for the renormalization factor $Z_5$. The $\as$-term in Eq.~(\ref{Z5}) is
required to restore the usual chiral Ward identity corresponding to the
conservation of the \oas\ axial current in the fermionic zero-mass limit.

With the $\g5$-prescription of Eq.~(\ref{rep}), we obtain the following
expression for the $\EuScript C$-odd part in the differential cross section
\be\label{vavirt}
L^{[\mu\nu]}\,H_{[\mu\nu]}^{V\!A+AV}(virtual\kern1pt) =
4(4-3\eps)\,q^2(1+\real{A}+\real{C})v\ct,
\ee
where the product of the two epsilon tensors has been replaced by the
following natural extension of the four-dimensional identity to $N$
dimensions
\be
\epsilon(\mu,\nu,\alpha,\beta) \epsilon(\mu,\nu,\rho,\sigma) \to
-2\Big(\,
\delta_{\alpha\rho}\delta_{\beta\sigma}-\delta_{\alpha\sigma}\delta_{\beta\rho}
\,\Big).
\ee
Note that the $\eps$-term in Eq.~(\ref{vavirt}) gives additional finite
contributions in the virtual part due to the infrared poles in $A$ and $C$
when compared with the result Eq.~(\ref{r2}) obtained by dimensional reduction.
\section{Real-Gluon Emission}
The \oas\ tree-graph contributions of Fig.~1(c) are required to cancel
the IR/M divergences in the virtual parts, and thus render the physical
cross section finite in concordance with the Kinoshita-Lee-Nauenberg
theorem~\cite{KLN}. In the following, we distinguish two types of IR/M
singularities that occur at one-loop level in the real and
virtual parts.

The {\em soft} divergences result from the massless character
of gluon field and are regulated by introducing a small gluon mass $m_g$.
The corresponding regulator $\Lambda=m_g^2/q^2$ is compatible with the cut-off
definition in the form factors Eqs.~(\ref{ff1}) and (\ref{ABC}),
and defines the scale to discriminate the soft-gluon from the hard-gluon
domain. In the case of real-gluon emission, $\Lambda\neq0$ has the specific task
to slightly deform the critical boundary region of the notorious three-body
phase-space. On the other hand, the {\em collinear} divergences emerge when the
gluon field couples with another massless field, and they typically manifest
themselves as singularities in the mass parameter $\xi=4m^2/q^2$.

Deriving the hadron tensors $h^{ij}_{\mu\nu}$ $(i,j=V,A)$ for the \oas\
tree-graph contributions, we only
have to consider the lowest-order results in the gluon-mass expansion since
the soft divergences, which will occur after performing the phase-space
integration, are at most logarithmical, and thus vanish in the limit
$\Lambda\to0$ when multiplied by an additional $\Lambda$. Four-dimensional
trace algebra gives the following explicit results for the parity-even
contributions
\bea
h^{VV}_{(\mu\nu)} &=&
8\pi\,\as\,\CF\,\Bigg[\,
g_{\mu\nu}\,\left\{{\xi\over y^2}+{4\over y}+{\xi\over z^2}+{4\over z}
-2{2-\xi\over yz}-2{y\over z}-2{z\over y}\right\} \nn
&&\kern2cm +4\,{p_{1\mu}p_{1\nu}\over q^2}\left\{{\xi\over y^2}-{2\over yz}
\right\}+4\,{p_{2\mu}p_{2\nu}\over q^2}\left\{{\xi\over z^2}-{2\over yz}
\right\} \nn
&&\kern2cm -4\,{p_{1\mu}p_{2\nu}+p_{1\nu}p_{2\mu}\over q^2}\,{\xi\over yz}
\,\Bigg], \label{hvvr} \\
h^{AA}_{(\mu\nu)} &=&
8\pi\,\as\,\CF\,\Bigg[\,
g_{\mu\nu}\,\left\{ 4{1-\xi\over y}+\xi{1-\xi\over y^2}+
                    4{1-\xi\over z}+\xi{1-\xi\over z^2} \right.\nn
&&\kern3cm \left.  -2{(1-\xi)(2-\xi)\over yz}-2{y\over z}-2{z\over y}
           \right\} \nn
&&\kern2cm +4\,{p_{1\mu}p_{1\nu}\over q^2}\left\{{\xi\over y^2}-2{1-\xi\over yz}
\right\}+4\,{p_{2\mu}p_{2\nu}\over q^2}\left\{{\xi\over z^2}-2{1-\xi\over yz}
\right\} \nn
&&\kern2cm +4\,{p_{1\mu}p_{2\nu}+p_{1\nu}p_{2\mu}\over q^2}\,{\xi\over yz}
\,\Bigg], \label{haar}
\eea
where the phase-space variables
\be\label{yz}
y = 1 - {2\,p_1\!\cdot q\over q^2}\qquad\mbox{and}\qquad
z = 1 - {2\,p_2\!\cdot q\over q^2}
\ee
specify the energies carried by quark and antiquark in the cms frame.
Energy-momentum conservation in combination with current conservation
eliminated the gluon momentum $p_3$ and lead to drastic simplifications
in these final expressions.

In the computation of the $\EuScript C$-odd contribution these favorable
features are absent. Especially the $\g5$-replacement prescription
Eq.~(\ref{rep}) yields quite lengthy expressions in the original trace
\bea\label{trace5}
h^{V\!A}_{[\mu\nu]} &=&
h^{AV}_{[\mu\nu]}(real\kern1pt) \nn
&=& -\Frac{1}{2}\;Tr (\Slash{p_1}+m)\,\Bigg[\,
\gamma_\alpha\,{\Slash{p_1}+\Slash{p_3}+m\over 2\,p_1\!\cdot p_3}\,
\gamma_{[\mu} +\gamma_{[\mu}\,{-\Slash{p_2}-\Slash{p_3}+m\over
2\,p_2\!\cdot p_3}\,\gamma_\alpha \,\Bigg] \\
&&\kern1.2cm (\Slash{p_2}-m)\,\Bigg[\,
\gamma^\alpha\,{-\Slash{p_2}-\Slash{p_3}+m\over 2\,p_2\!\cdot p_3}\,
\gamma_{\nu]}\g5+\gamma_{\nu]}\g5\,{\Slash{p_1}+\Slash{p_3}+m\over
2\,p_1\!\cdot p_3}\,\gamma^\alpha\,\Bigg]. \nonumber
\eea
At this point, we shall therefore not reproduce the full expressions for the
uncontracted hadron tensor $h^{V\!A}_{[\mu\nu]}$ but postpone further
discussion of the $\EuScript C$-odd part until we will come to the final
contracted results.

The $q\bar{q}g$ phase-space is considerably more complicated than the
previously discussed two-particle complement. In general, five parameters
are needed to completely characterize a specific configuration in
three-particle phase-space. The phase-space parametrization is crucial
for obtaining closed analytical solutions after integration out all variables
except the $\ct$-dependence and the cms energy. Our particular choice for these
five parameters is $\sqrt{q^2}$, $y$, $z$, $\ct$, and $\varphi$.

Energy-momentum conservation confines the momenta of the three outgoing
particles to within a plane. Fig.~3 illustrates the angular orientation of
the $({\bm p_1}, {\bm p_2}, {\bm p_3})$-plane in the cms coordinate frame.
The vector normal to the outgoing 3-jet plane is
\be
\ndirec = \Big(-\ct\,\cos\varphi,\,-\ct\,\sin\varphi,\,\st\,\Big),
\ee
and the direction of the scattered antiquark ${\bm{\hat{p}_2}}$
is obtained by a rotation $\bm{\EuScript R}$ of $\direc={\bm p_1}/|{\bm p_1}|$
around $\ndirec$ with the angle $\chi$. We find
\be\label{p2u}
\bm{\hat{p}_2} = \bm{{\EuScript R}_{\,\hat{n}}}(\chi)\,\direc =
\direc\,\cos\chi-\ndirec\times\direc\,\sin\chi.
\ee
Now it is straightforward to write all final momenta in terms of the
chosen phase-space parameters (with the standard metric
$g^{\mu\nu}=\mbox{diag}(1;-1,-1,-1)$)
\bea
p_1 & =&
\Frac{1}{2}\sqrt{q^2}\,\Big(1-y;\;\direc\,\sqrt{(1-y)^2-\xi}\,\Big), \nn
p_2 & =&
\Frac{1}{2}\sqrt{q^2}\,\Big(1-z;\;\bm{\hat{p}_2}\,\sqrt{(1-z)^2-\xi}\,\Big),\\
p_3 & =&
\Frac{1}{2}\sqrt{q^2}\,(y+z)\,\Big(\,1;\;\bm{\hat{p}_3}\,\Big), \nonumber
\eea
where the gluon is scattered off in the direction
\be
\bm{\hat{p}_3} =
-\direc\,{\sqrt{(1-y)^2-\xi}\over y+z}-
\bm{\hat{p}_2}{\,\sqrt{(1-z)^2-\xi}\over y+z}.
\ee
Note that $p_3^2=0$ automatically requires $\bm{\hat{p}_3}$ to have unit
length which yields in combination with Eq.~(\ref{p2u})
\be\label{cx}
\cos\chi = \direc\cdot\bm{\hat{p}_2} =
{y+z+y z+\xi-1\over\sqrt{(1-y)^2-\xi}\sqrt{(1-z)^2-\xi}},
\ee
which makes evident that $\chi$ gives no additional degree of freedom but
is fully determined by energy-momentum conservation.
It is now easy to obtain the following relations to classify the angular
dependence that the various hadronic components of Eqs.~(\ref{hvvr}) and
(\ref{haar}) give upon contraction with the lepton tensor
\bea
L^{\mu\nu}\,g_{\mu\nu} &=& -4, \\
L^{\mu\nu}\,{p_{1\mu}p_{1\nu}\over q^2} &=&
\Frac{1}{2}\s2t\,\Big\{\,(1-y)^2-\xi\,\Big\}, \\
L^{\mu\nu}\,{p_{2\mu}p_{2\nu}\over q^2} &=&
\Frac{1}{2}\left(1-\cos^2\chi\,\c2t\right)\,\Big\{\,(1-z)^2-\xi\,\Big\}, \\
L^{\mu\nu}\,{p_{1\mu}p_{2\nu}+p_{1\nu}p_{2\mu}\over q^2} &=&
\Frac{1}{2}\s2t\,\Big\{\,y+z+y z+\xi-1\,\Big\}.
\eea
Hence, we readily find the following contracted results after substituting
for $\cos\chi$ according to Eq.~(\ref{cx})
\bea
&& L^{\mu\nu}\,h^{VV}_{\mu\nu} = 8\pi\,\as\,\CF\,\Bigg[ \nn
&&-8(1+\xi)\left({1\over y}+{1\over z}\right)-2\xi(1+\xi)\left(
{1\over y^2}+{1\over z^2}\right)
+4\left({y\over z}+{z\over y}\right)+4(1+\xi)(2-\xi){1\over yz} \nn
&&+{2\ct\over(1-y)^2-\xi}\,\Bigg\{
-5\xi-2(1-\xi)(2-3\xi){1\over y}-\xi(1-\xi)^2{1\over y^2}
-4(1-\xi)(3-2\xi){1\over z} \nn
&&\kern2.65cm-\xi(1-\xi)^2{1\over z^2}
-2\xi(1-\xi){y\over z^2}-\xi{y^2\over z^2}+4z+2{z\over y}
+2(7-5\xi){y\over z} \nn
&&\kern2.65cm+8{y^2\over z}+2(2-\xi)(1-\xi)^2{1\over yz}+2{y^3\over z}+4y+2yz
\Bigg\}\quad\Bigg], \\
&&\mbox{and} \nn
&& L^{\mu\nu}\,h^{AA}_{\mu\nu} = 8\pi\,\as\,\CF\,\Bigg[ \nn
&&8\xi-8(1-\xi))\left({1\over y}+{1\over z}\right)-2\xi(1-\xi)
\left({1\over y^2}+{1\over z^2}\right)
+4(1+\xi)\left({y\over z}+{z\over y}\right) \nn
&&+4(1-\xi)(2-\xi){1\over yz}
+{2\ct\over(1-y)^2-\xi}\,\Bigg\{
-\xi-2(1-\xi)(2-3\xi){1\over y}-\xi(1-\xi)^2{1\over y^2} \nn
&&-4(1-\xi)(3-2\xi){1\over z}-\xi(1-\xi)^2{1\over z^2}
+2\xi(1-\xi){y\over z^2}+2(1-\xi){y^3\over z}
-\xi{y^2\over z^2}+4(1-\xi)z \nn
&&+2(1-\xi){z\over y}+2(7-6\xi){y\over z}-4(2-\xi){y^2\over z}
+2(2-\xi)(1-\xi)^2{1\over yz}+4y-4\xi y^2 \nn
&&+2(1-\xi)y z
\Bigg\}\quad\Bigg].
\eea
Out of the five variables that parametrize the three-particle phase-space
we only need to remove the $y$- and $z$-dependence by integration to yield
the differential cross section, depending solely on $E_{cms}=\sqrt{q^2}$
and $\ct$. Note that in Eq.~(\ref{master}) the $\varphi$-integration has
already been included in the two-particle phase-space factor PS$_2$ apart
from the flux factor. Thus, the appropriate $(y,z)$-integration can be put
into the form
\be\label{rpsf}
H^{ij}_{\mu\nu}(real\kern1pt)
= \int{d^4 \mbox{PS}_3\over\mbox{PS}_2}\:h^{ij}_{\mu\nu} =
{q^2\over 16\pi^2\,v}\int\limits_{y_-}^{y_+}\!dy\int\limits_{z_-(y)}^{z_+(y)}
\!\!dz\:h^{ij}_{\mu\nu},
\ee
which gives the final real-gluon contributions that have to be included in
Eq.~(\ref{master}) for the cancellation of all IR divergences with the
corresponding virtual parts.

The complicated upper and lower bounds of the nested integral in
Eq.~(\ref{rpsf}) are given in Appendix~B where we discuss in more detail
the $q\bar{q}g$ phase-space.
In our scheme, the phase-space integrals are grouped into distinct classes
$\{\I{i}\}, \{\S{i}\}$, and $\{\J{i}\}$. Each class refers to a different
functional dependence on the quark's velocity in the three-body system.
The individual elements of $\{\I{i}\}, \{\S{i}\}$, and $\{\J{i}\}$ are
identified in Appendix~B which also treats some specific properties of the
integrals and their interconnection.

Thus, the final integrated results for the \oas\ tree graphs with $VV$ and $AA$
parity-parity combinations are
\bea
&&
L^{\mu\nu}\,H^{VV}_{\mu\nu}(real\kern1pt) =
{q^2\over2v}\,{\as\over\pi}\,\CF\,\Bigg[\,
-16(1+\xi)\I2-4\xi(1+\xi)\I3+8\I4+4(1+\xi)(2-\xi)\I5 \nn[.3cm]
&&+2\c2t\,\Bigg\{\,-8\I2-5\xi\J1-2(1-\xi)(2-3\xi)\J2-\xi(1-\xi)^2\J3
+4\J4+2\J6+4\J7 \nn[.3cm]
&&-4(1-\xi)(1-2\xi)\J8-\xi(1-\xi)^2\J9
+2(2-\xi)(1-\xi)^2\J{10}+2\J{11}-2(1+5\xi)\J{12} \nn[.3cm]
&&+2\xi(1-\xi)\J{13}-\xi\J{14}+2\J{15}
\,\Bigg\}\quad\Bigg], \\[.3cm]
&&
L^{\mu\nu}\,H^{AA}_{\mu\nu}(real\kern1pt) =
{q^2\over2v}\,{\as\over\pi}\,\CF\,\Bigg[\,
8\xi\I1-16(1-\xi)\I2-4\xi(1-\xi)\I3+8(1+\xi)\I4 \nn[.3cm]
&&+4(1-\xi)(2-\xi)\I5+2\c2t\,\Bigg\{\,
-4(2-\xi)\I2-\xi\J1-2(1-\xi)(2-3\xi)\J2-\xi(1-\xi)^2\J3 \nn[.3cm]
&&+4\J4-4\xi\J5+2(1-\xi)\J6+4(1-\xi)\J7-4(1-\xi)^2\J8-\xi(1-\xi)^2\J9
+2(2-\xi)(1-\xi)^2\J{10} \nn[.3cm]
&&+2(1-\xi)\J{11}-2(1+2\xi)\J{12}+2\xi(1-\xi)\J{13}
-\xi\J{14}+2(1-\xi)\J{15}
\,\Bigg\}\quad\Bigg].
\eea
Note that $\{\I{i}\}, \{\S{i}\}$, and $\{\J{i}\}$ are the smallest units
of the three-particle phase-space one arrives at after parametrization
in the energy variables $(y,z)$ and subsequent partial fractioning. These
single components are process-independent. With the additional integral
class $\T{i}$ (see Ref.~\cite{beam}) they constitute a complete set of
solutions for any one-loop real-emission process from massive fermions.
Differential or total observables such as production rates or various
polarization components can all be expressed succinctly in terms of these
units once they are derived.

To contract the asymmetric part of the lepton tensor $L^{[\mu\nu]}$ with
the $\g5$-odd hadronic complement Eq.~(\ref{trace5}), we use the following
identities
\bea
&&L^{\mu\nu}\,{\epsilon(\mu,\nu,p_1,q)\over q^2} \ =\
i\,2v\,\ct, \label{first} \\
&&L^{\mu\nu}\,{\epsilon(\mu,\nu,p_2,q)\over q^2} \ =\
-i\,2v\,\ct, \\
&&L^{\mu\nu}\,{\epsilon(\mu,\nu,p_1,p_2)\over q^2} \ =\
-i\,{2\ct\over\sqrt{(1-y)^2-\xi}}\,\Big\{\,
(2-\Frac{1}{2}\xi)y+(1-\Frac{1}{2}\xi)z \nn
&&\kern8cm -y^2-y z+\xi-1\,\Big\}, \\
&&L^{\mu\nu}\,
{\epsilon(\mu,p_1,p_2,q)\,p_{2\nu}-\epsilon(\nu,p_1,p_2,q)\,p_{2\mu}\over
(q^2)^2} \ =\ i\,v\,\ct\,y, \\
&&L^{\mu\nu}\,
{\epsilon(\mu,p_1,p_2,q)\,q_\nu-\epsilon(\nu,p_1,p_2,q)\,q_\mu\over
(q^2)^2} \ =\ i\,v\,\ct\,(y+z). \label{last}
\eea
In these expressions, the contraction of two epsilon tensors with one common
index has been replaced by an appropriately antisymmetrized product of metric
tensors which is well-defined in either four or $N$ dimensions:
\be\label{epseps}
\epsilon_{\mu\alpha\beta\gamma}\,\epsilon^{\mu\rho\sigma\tau} =
-3!\,g_\alpha^{\ [\rho}\,g_\beta^{\ \sigma}\,g_\gamma^{\ \tau]}.
\ee
Then, the scalar product of two Lorentz vectors is invariant in different
space-time dimensions so that Eqs.~(\ref{first})--(\ref{last}) give unique
results.

It is now straightforward to obtain the following contracted results for the
general case of $N$-dimensional Clifford algebra
\bea\label{lh5}
&&
L^{\mu\nu}\,h^{V\!A}_{\mu\nu} =
L^{\mu\nu}\,h^{AV}_{\mu\nu} \nn[.3cm]
&&=\ 32\pi\,\as\,\CF\,\ct\,
\left(1-\Frac{3}{4}\eps\right)\Bigg[\,
\eps\xi-(4-5\xi){1\over y}
-\xi(1-\xi)\left({1\over y^2}+{1\over z^2}\right)
\nn[.3cm] &&\kern.4cm
-2(4-3\xi){1\over z}+\xi{y\over z^2}+2\eps y
+(2+\eps)z+\left\{2-\eps\left(1-\Frac{1}{2}\xi\right)\right\}{z\over y}
\\[.3cm] &&\kern.4cm
+\left\{6-\eps\left(1-\Frac{1}{2}\xi\right)\right\}{y\over z}
-(2-\eps){y^2\over z}+(1-\xi)(2-\xi){1\over yz}
+{\EuScript O(\eps^2)}\,\Bigg]. \nonumber
\eea

Putting $\eps$ in Eq.~(\ref{lh5}) to zero and then performing the
$(y,z)$-integration gives the following result for the $\EuScript C$-odd
tree-graph contribution within dimensional reduction
\bea\label{dr5}
&&
L^{\mu\nu}\,H^{V\!A}_{\mu\nu}(real\kern1pt) =
L^{\mu\nu}\,H^{AV}_{\mu\nu}(real\kern1pt)
\nn[.3cm] &&
=\ 2\,{q^2\over v}\,{\as\over\pi}\,\CF\,\ct\,\Bigg[\,
-(4-5\xi)\S2-\xi(1-\xi)(\S3+\S5)-2(4-3\xi)\S4
\nn[.3cm] &&\kern.4cm
+\xi\S6+2\S8+2\S9+6\S{10}-2\S{11}+2(1-\xi)(2-\xi)\S{12}
\,\Bigg]. \label{vared}
\eea
On the other hand, adopting the $\g5$-prescription of Eq.~(\ref{rep})
we obtain
\bea
&&
L^{\mu\nu}\,H^{V\!A}_{\mu\nu}(real\kern1pt) =
L^{\mu\nu}\,H^{AV}_{\mu\nu}(real\kern1pt)
\nn[.3cm] &&
=\ 2\,{q^2\over v}\,{\as\over\pi}\,\CF\,\ct\,
\left(1-\Frac{3}{4}\eps\right)\Bigg[\,
\eps\xi\S1-(4-5\xi)\S2-\xi(1-\xi)(\S3+\S5)-2(4-3\xi)\S4
\nn[.3cm] &&\kern.4cm
+\xi\S6+(2+\eps)\S8+
\left\{2-\eps\left(1-\Frac{1}{2}\xi\right)\right\}\S9
+\left\{6-\eps\left(1-\Frac{1}{2}\xi\right)\right\}\S{10}
\label{va1}\\[.3cm] &&\kern.4cm
-(2-\eps)\S{11}+2(1-\xi)(2-\xi)\S{12}+2\eps\S{13}
+{\EuScript O(\eps^2)}\,\Bigg]
\nn[.3cm] &&
=\ 2\,{q^2\over v}\,{\as\over\pi}\,\CF\,\ct\,\Bigg[\,
-(4-5\xi)\S2-(1-\Frac{3}{4}\eps)\xi(1-\xi)(\S3+\S5)-2(4-3\xi)\S4
\nn[.3cm] &&\kern.4cm
+\xi\S6+2\S8+2\S9+6\S{10}-2\S{11}+2(1-\Frac{3}{4}\eps)(1-\xi)(2-\xi)\S{12}
+{\EuScript O(\eps)}\,\Bigg]. \label{va2}
\eea
Eq.~(\ref{va1}) displays the full result neglecting $\EuScript O(\eps^2)$
bracket terms, which do not contribute in the limit $\eps\to0$. Since the
logarithmic divergences within the massive gluon scheme correspond at \oas\
to the $1/\eps$ poles of a dimensionally regularized theory, we can not naively
discard all $\EuScript O(\eps)$-terms. In Eq.~(\ref{va2}) only the non-vanishing
$\EuScript O(\eps)$ bracket terms are retained.
However, these additional $\EuScript O(\eps)$ terms represent the finite
difference in the $\g5$-odd real-gluon correction between dimensional reduction
and dimensional regularization with the $\g5$-replacement scheme. Finally,
when the corresponding virtual-gluon correction Eq.~(\ref{vavirt}) is added,
these contributions disappear, and one obtains the same total result as in
dimensional reduction Eq.~(\ref{dr5}).
\section{Complete {\boldmath\oas} Differential Cross Sections}
With the explicit expressions for the virtual and real contributions of the
previous sections, we are now in the position to give the full analytical
results for the \oas\ differential cross section of the annihilation
process $e^+ e^-\to\gamma,Z\to q\bar{q}$. Full quark-mass dependence was
kept and no other approximations were made so that these analytical formulae
are valid in a closed form over the entire energy range of perturbative QCD.
All angular dependence resulted naturally from basic kinematics of the two- and
three-body final state.

A severe test on the total results is provided by the cancellation of the
soft and collinear divergences separately for each angular distribution
within the corresponding real and virtual parity-parity combinations. Using
the relative phase-space factor Eq.~(\ref{rpsf}) in the sum of virtual and
real gluon parts, we obtain the following total \oas\ results (including the
Born level)
\bea
&& L^{\mu\nu}\,H^{VV}_{\mu\nu}(total\kern1pt) =4q^2\,\Bigg[\,
(1+2\real{\tilde{A}})\left\{2-v^2(1-\c2t)\right\}+2\real{B}\,v^2(1-\c2t)
\nn[.2cm]
&&+{\as\over4\pi}\,\CF\,{1\over v}\,\Bigg\{\,
-8(1+\xi)\I2-2\xi(1+\xi)\II3+4\I4+2(1+\xi)(2-\xi)\II5 \nn[.3cm]
&&+\c2t\,\Big\{-8\I2-5\xi\J1-2(1-\xi)(2-3\xi)\J2-\xi(1-\xi)^2\JJ3+4\J4+2\J6
+4\J7 \nn[.3cm]
&&-4(1-\xi)(1-2\xi)\J8-\xi(1-\xi)^2\JJ9+2(2-\xi)(1-\xi)^2\JJ{10}+2\J{11}
-2(1+5\xi)\J{12} \nn[.3cm]
&&+2\xi(1-\xi)\J{13}-\xi\J{14}+2\J{15}\Big\}\,
\Bigg\}\,\Bigg], \label{hvvtot} \\[.5cm]
&& L^{\mu\nu}\,H^{AA}_{\mu\nu}(total\kern1pt) =4q^2\,\Bigg[\,
(1+2\real{\tilde{C}})\,v^2(1+\c2t) \nn[.2cm]
&&+{\as\over4\pi}\,\CF\,{1\over v}\,\Bigg\{\,
4\xi\I1-8(1-\xi)\I2-2\xi(1-\xi)\II3+4(1+\xi)\I4+2(1-\xi)(2-\xi)\II5 \nn[.3cm]
&&+\c2t\,\Big\{-4(2-\xi)\I2-\xi\J1-2(1-\xi)(2-3\xi)\J2-\xi(1-\xi)^2\JJ3+4\J4
-4\xi\J5 \nn[.3cm]
&&+2(1-\xi)\J6+4(1-\xi)\J7-4(1-\xi)^2\J8-\xi(1-\xi)^2\JJ9
+2(2-\xi)(1-\xi)^2\JJ{10} \nn[.3cm]
&&+2(1-\xi)\J{11}-2(1+2\xi)\J{12}+2\xi(1-\xi)\J{13}-\xi\J{14}
+2(1-\xi)\J{15}\Big\}\,
\Bigg\}\,\Bigg],  \label{haatot}
\eea
where the wiggle on top of form factors or integrals denotes quantities free
from soft divergences as $\Lambda\to0$, which essentially means that only
linear terms in $\ln\Lambda$ have been discarded. We find that in the massless 
limit $\xi\to0$ all collinear singularities cancel, as expected. It is now
easy to obtain the full \oas\ expressions for the structure functions in
Eq.~(\ref{1}) by using the convolutions
\be
\sigma_{\scriptscriptstyle U,L} =
{3\over 8}\pi\left({\alpha\over q^2}\right)^2 v\,\sum_{ij=VV,AA}
g^{ij}\:\Pi_{\scriptscriptstyle U,L}\:L^{(\mu\nu)}\,H_{(\mu\nu)}^{ij},
\ee
where the unpolarized and longitudinal projectors are explicitly given by
\bea
\Pi_{\scriptscriptstyle U} &=& \int\limits_{-1}^{+1}\!d\ct\;
\Big\{\,5\c2t-1\,\Big\}, \\
\Pi_{\scriptscriptstyle L} &=& \int\limits_{-1}^{+1}\!d\ct\;
\Big\{\,2-5\c2t\,\Big\}.
\eea

Similarly, one finds that the remaining total $V\!A$ combination is IR finite.
Adding Eqs.~(\ref{vavirt}) and (\ref{va2}) gives for the $\EuScript C$-odd
component of the differential rate
\bea\label{fb}
&& \sf = 8\pi\,{\alpha^2\over q^2}\,v\,g^{V\!A}\Bigg[\,
(1+\real{\tilde{A}}+\real{\tilde{C}})\,v \nn[.2cm]
&&+{\as\over4\pi}\,\CF\,{1\over v}\,\Bigg\{\,
-(4-5\xi)\S2-\xi(1-\xi)(\SS3+\SS5)-2(4-3\xi)\S4+\xi\S6 \nn[.3cm]
&&+2(\S8+\S9+3\S{10}-\S{11})+2(1-\xi)(2-\xi)\SS{12}\Big\}\,
\Bigg],
\eea
where we substituted $A$ and $C$ according to Eqs.~(\ref{ff1}) and (\ref{C5})
without soft divergences. Note that for the $N$-dimensional
$\g5$-replacement scheme the terms proportional to $\eps$ in the real part
have exactly canceled with the extra contributions in the virtual part. Thus,
dimensional reduction and dimensional regularization give the same result for
the forward-backward asymmetry in the differential cross section. The
independent result for dimensional reduction is easily obtained by adding
Eqs.~(\ref{red1}) and (\ref{vared}).

In the massless calculation, there are no QCD one-loop contributions to the
forward-backward asymmetry~\cite{JLZ,bodo}. Apart from the cancellation of
the spurious IR divergences in Eq.~(\ref{fb}), we recover this specific result
in the limit $\xi\to0$. On the other hand, we have checked explictly that
integrating the $VV$ and $AA$ parts of the differential rate over $\ct$
reproduces exactly the analytical results for the total cross section as
found in Ref.~\cite{grun}.

The numerical estimates of the \oas\ differential cross section for bottom
quark production with a fixed-point mass $m_b(m_b)=4.3\,$ GeV are shown in
Fig.~4. The running of the strong coupling is implemented by taking
$\as^{(5)}(\mz)=0.123$ in the modified minimal subtraction scheme
$(\overline{\mbox{\rm MS}})$ for five active flavors. For the bottom quark
at energies above the next flavor threshold and in general for the top quark,
we apply the appropriate matching conditions for six active flavors
using the corresponding one-loop QCD renormalization group equations~\cite{run}.

Fig.~4a gives a surface plot of the \oas\ differential rate as a function
of the cms energy $E_{cms}=\sqrt{q^2}$ and the cosine of the scattering angle
$\ct$. On the $Z$-peak, the asymmetry is clearly pronounced and yields for
$E_{cms}=\mz=91.178\,$ GeV values ranging from 5800 pb $(\theta=\pi)$ to
8300 pb $(\theta=0)$. The minimum is located at $\ct\approx0.16$ with 3494 pb.
For higher energies the differential rate falls rapidly off
to give at 100 GeV cross sections of the order of only 100 pb. Nevertheless,
the massive \oas\ corrections become increasingly important off the $Z$-peak.
In Fig.~4b, the \oas\ results are compared with the Born approximation.
Note that the dominant \oas\ contributions are below the $Z$-peak in
the domain $0<\ct<0.4$ and above the $Z$-peak in the domain $-0.5<\ct<0$.
Thus, the area of maximum correction (4.3--4.5\,\%) consists of a strip of
approximately $\pi/6$ width that shifts from the upper to the lower
hemisphere when passing the $Z$-threshold (see also Fig.~2). At 100 GeV the
corrections amount for $\ct=0.3$ to full 4.5\,\%.

Fig.~5a depicts a similar three-dimensional plot with the differential cross
section for $e^+ e^-\to\gamma,Z\to t\bar{t}$ at one-loop QCD level. The
value for the top mass is $m_t(m_t)=174\,$GeV. Note that we have chosen an
energy range sufficiently higher than the $t\bar{t}$ threshold to avoid
full interference with the non-perturbative sector. The dominant contribution
to the differential rate is given by top quarks scattered in the forward
direction. Along the forward direction, we find at $\sim425\,$GeV a peak
value of $\sim0.85$~pb for the cross section which includes an \oas\
correction of nearly 20\,\%. At 350 GeV the strong corrections have an
impact of 52\,\%. Thus, it is conceivable that for top quark production
non-perturbative effects still prevail in energy domains considerably above
the $t\bar{t}$ threshold.

In Fig.~5b, we give the distribution of the energy and scattering regions most
significant to the strong corrections. Note that the \oas\ terms contribute
dominantly along the forward-scattering axis where the differential rate
takes maximum values. On the other hand, bottom production yielded the most
important \oas\ corrections perpendicular to the beam axis which corresponds
to minimal values on the saddle surface in Fig.~4a.
\section{Conclusions}
In the present work, we have derived from first principles the full analytical
\oas\ results for the differential cross sections in heavy-quark production.
No mass approximations or further restrictive assumptions were made.
All angular distributions arose naturally from the underlying phase-space
kinematics.

The systematic treatment of the tree-graph contributions relies on the exact
integral solutions of the massive three-body phase-space. These integrals
are divided into several classes according to their functional dependence
on the final quark velocity. A complete collection of these phase-space
integrals allows to describe any one-loop bremsstrahlung process including
mass effects.

In particular, we put emphasis on the consistent treatment of the axial-vector
current to handle the spurious chiral anomalies that are usually present
in dimensionally regularized calculations with an odd number of $\g5$'s.
We explicitly showed that dimensional reduction and dimensional regularization
with the $\g5$-replacement prescription yield identical results for the
forward-backward asymmetry.

In our numerical estimates, we found that at one-loop level the final state
QCD corrections already modify the Born approximation by approximately
4.5\,\% (100 GeV) for bottom production and more than 25\,\% (375 GeV) for top
production. We presented details on the differential distribution, and made
the essential observation that for the bottom quark the \oas\ corrections
are dominant in the scattering region perpendicular to the beam axis
whereas for top quarks \oas\ contributions become very important along
the forward-scattering axis.

We conclude that full analytical calculations for heavy-fermion pair production
at one-loop order are feasible and provide an attractive alternative to
existing Monte-Carlo generators.
\vskip1cm\noindent
{\bf Acknowledgements.} M.M.T.\ wishes to thank Sergei Larin for
clarifying comments on the treatment of the axial-vector current
within dimensional regularization, and Sergio Novaes for a stimulating
discussion on the numerical results. He further gratefully acknowledges
the support given by the Alexander-von-Humboldt Foundation. This work
has been supported by CICYT under grant AEN93-0234 and IVEI under
grant 03-007.

\newpage
\section*{Appendix A:\hskip4mm Renormalization of the Axial-Vector
          \hphantom{Appendix A:}\hskip7mm Current}
\setcounter{equation}{0}
\def\theequation{A\arabic{equation}}
In this appendix, we give a brief account of the subtleties associated with
the renormalization of the axial-vector current at QCD one-loop level.

As we explained in Section~2, dimensional regularization with a conventional
anticommuting $\g5$ or a replaced $\g5$ produces identical results for
the gluonic vertex correction to the axial-vector current. Explicitly, we
gave meaning to the $\g5$-matrix in a dimension other than four by the
replacement rule
$$
\gamma_\mu\g5\,\to\,Z_5\,{i\over 3!}\,
\epsilon_\mu^{\hphantom{\mu}\rho\sigma\tau}\,
\gamma_\rho\gamma_\sigma\gamma_\tau,
$$
where the finite renormalization constant $Z_5$ is necessary to reinstate the
validity of the axial-vector Ward identities in the final
results~\cite{larin,LV}.

Using Eq.~(\ref{epseps}) one can recast Eq.~(\ref{C1}) into the following
expression for the unrenormalized form factor
\bea\label{Cbare}
C_{\mbox{\tiny bare}} &=& -8{\pi i\over 3N-8}\,\as\CF
{g^{\nu\,[\rho}\,p_1^\sigma\,p_2^{\tau\,]}\over v^2(q^2)^2} \\
    & &
  \times\kern5pt\int{d^N k\over(2\pi)^N}
  {\Tr (\Slash{p_1}+m)\,\gamma_\alpha\,(\Slash{p_1}+\Slash{k}+m)\,
  \gamma_\rho\gamma_\sigma\gamma_\tau\,
  (-\Slash{p_2}+\Slash{k}+m)\,\gamma^\alpha\,
  (\Slash{p_2}-m)\,\gamma_\nu \over
  k^2\,\Big\{(p_1+k)^2-m^2\Big\}\,\Big\{(p_2-k)^2-m^2\Big\} },
  \nonumber
\eea
where all indices are $N$-dimensional. Note that the normalization has to
be in agreement with the overall $N$-dependent factor in the Born contribution
Eq.~(\ref{vavirt}).

In the solution of this one-loop integral we use the following structure for
the two- and three-point functions ($p_1^2=p_2^2=m^2$)
\bea
I_3(p_1,p_2) &=& \int{d^N k\over(2\pi)^N}
{ 1\over k^2\,\Big\{(p_1+k)^2-m^2\Big\}\,\Big\{(p_2-k)^2-m^2\Big\} }
= {P(v)\over m^2}, \label{ifirst} \\[.3cm]
I_3^\mu(p_1,p_2) &=& \int{d^N k\over(2\pi)^N}
{ k^\mu\over k^2\,\Big\{(p_1+k)^2-m^2\Big\}\,\Big\{(p_2-k)^2-m^2\Big\} }
\nn &=& Q(v){(p_1-p_2)^\mu\over m^2}, \\[.3cm]
I_3^{\mu\nu}(p_1,p_2) &=& \int{d^N k\over(2\pi)^N}
{ k^\mu k^\nu\over k^2\,\Big\{(p_1+k)^2-m^2\Big\}\,\Big\{(p_2-k)^2-m^2\Big\} }
\\ &=& R(v)\,g^{\mu\nu} + S(v) {(p_1-p_2)^\mu(p_1-p_2)^\nu\over m^2} +
  T(v) {(p_1+p_2)^\mu(p_1+p_2)^\nu\over q^2}, \nn[.3cm]
I_2(p_1,p_2) &=& \int{d^N k\over(2\pi)^N}
{ 1\over \Big\{(p_1+k)^2-m^2\Big\}\,\Big\{(p_2-k)^2-m^2\Big\} }
= g_{\mu\nu} I_3^{\mu\nu}(p_1,p_2). \label{ilast}
\eea
Notice that the last identity only holds if one takes
$g^{\mu\nu}\,g_{\mu\nu} = N = 4-\eps$. All coefficient functions
$P,\ldots,T$ are dimensionless and depend only on the mass parameter $v$.

An explicit calculation yields for Eq.~(\ref{Cbare})
\bea\label{Cbare2}
C_{\mbox{\tiny bare}} &=& 16\pi i\,\as\CF\,\left(1+\Frac{3}{4}\eps\right)
\Bigg[\,
\left(1-\Frac{3}{4}\eps\right){1+v^2\over1-v^2}\Big\{P(v)+2Q(v)\Big\}
-\left(1+\Frac{1}{4}\eps\right) R(v) \nn && \kern3.75cm
+\left(1-\Frac{1}{4}\eps\right){2v^2\over1-v^2} S(v)-\Frac{1}{8}(4-\eps) T(v)
+\EuScript O(\eps^2)\,\Bigg].
\eea
Here, the $\eps$-dependence stems exclusively from the $N$-dimensional trace
algebra so that the result for dimensional reduction is obtained from the above
formula by putting $\eps=0$. We verified this result also in an independent
calculation.

Adopting the conventions of Refs.~\cite{thooft,PV}, we can further decompose
the elements of Eqs.~(\ref{ifirst})--(\ref{ilast}) into the so-called
Passarino-Veltman functions
\bea
P(v)&=&{i\mu^{-\eps}\over(4\pi)^2}m^2\,C_0\left(m^2,q^2,m^2;0,m^2,m^2\right),\\
Q(v)&=&{i\mu^{-\eps}\over(4\pi)^2}{1-v^2\over4v^2}\Bigg[\,
B_0\left(q^2;m^2,m^2\right)-B_0\left(m^2;0,m^2\right)\,\Bigg], \\
R(v) &=& {i\mu^{-\eps}\over4(4\pi)^2}{1-v^2\over4v^2}\Bigg[\,
1+B_0\left(q^2;m^2,m^2\right)\,\Bigg], \\
S(v) &=& {i\mu^{-\eps}\over(4\pi)^2}{1-v^2\over16v^2}\Bigg[\,
B_0\left(m^2;0,m^2\right)-B_0\left(q^2;m^2,m^2\right)\,\Bigg], \\
T(v) &=& {i\mu^{-\eps}\over4(4\pi)^2}\Bigg[\,
B_0\left(m^2;0,m^2\right)-B_0\left(q^2;m^2,m^2\right)-2\,\Bigg].
\eea
For conciseness, we do not consider the imaginary contributions to the
form factor but concentrate on the real parts of the following general
scalar integral representations
\bea
&& B_0\left(\,p_1^2;m_0^2,m_1^2\,\right) =
   (i\,\pi^2)^{-1}\int d^N k\;{ 1\over
\Big\{k^2-m_0^2\Big\}\Big\{(k+p_1)^2-m_1^2\Big\} }, \\[.3cm]
&& C_0\left(\,p_1^2,(p_1-p_2)^2,p_2^2;m_0^2,m_1^2,m_2^2\,\right) = \nn
&& (i\,\pi^2)^{-1} \int d^N k\;{ 1\over
\Big\{k^2-m_0^2\Big\}\Big\{(k+p_1)^2-m_1^2\Big\}\,\Big\{(k+p_2)^2-m_2^2\Big\} },
\eea

In the required Passarino-Veltman functions we neglect terms proportional
to $\eps$ and drop the imaginary parts. These scalar two- and three-point
functions are divergent as $\eps\to0$ and read
\bea
B_0\left(m^2;0,m^2\right) &=&
{2\over\eps}-\gE+\ln\left({4\pi\mu^2\over m^2}\right)+2, \\
B_0\left(q^2;m^2,m^2\right) &=&
{2\over\eps}-\gE+\ln\left({4\pi\mu^2\over m^2}\right)+2+
v\,\ln\left({1-v\over1+v}\right), \\
C_0\left(m^2,q^2,m^2;0,m^2,m^2\right) &=&
{1\over q^2\,v}\Bigg[\,\left\{{2\over\eps}-\gE
+\ln\left({4\pi\mu^2\over m^2}\right)-{3v^2\over1+v^2}\right\}\,
\ln\left({1-v\over1+v}\right) \nn
&&\kern1.3cm-{v\over1+v^2}\Big\{F(v)+4\Big\}\,\Bigg],
\eea
where $F(v)$ was previously defined in Eq.~(\ref{F}). Here, we have included
the 't Hooft mass $\mu$ which naturally arises when the couplings are extended
to $N$ dimensions.

Thus, the additional real contribution induced by $\eps$ in Eq.~(\ref{Cbare2})
is given by
\be
\delta C := C_{\mbox{\tiny bare}}-C =
-16\pi i\,\as\CF\,\eps\,R(v)\,\to\,{\as\over2\pi}\CF
\ee
Note that the first term in brackets on the right-hand side depends on
the mass parameter $v$ and is thus connected with the infrared structure
of the theory. On the other hand, the remaining term stems from the
ultraviolet sector.

The UV divergences in Eq.~(\ref{Cbare}) are removed by renormalization.
Apart from the conventional on-shell renormalization scheme we take into
account the renormalization of the axial-vector current particular to
the special $\g5$-prescription used~\cite{larin}. Therefore, the entire
renormalization program amounts to isolating $C_{\mbox{\tiny rep}}$ in
the following equality
\be
1+C_{\mbox{\tiny rep}} = Z_5\,Z_2\,\Big(\,1+C_{\mbox{\tiny bare}}\,\Big),
\ee
where the subscript refers to the particular $\g5$-definition chosen. In the
$\overline{\mbox{\rm MS}}$ scheme, the quark-field renormalization constant
$Z_2$ (with gluon mass regularization) and the finite axial-vector
renormalization constant $Z_5$ at one-loop order are given by
\bea
Z_2 &=& 1+{\as\over4\pi}\,\CF\,\Bigg[\,-{2\over\eps}+\gE-
\ln\left({4\pi\mu^2\over m^2}\right)-4-2\ln\Lambda+2\ln\xi-4\ln2\,\Bigg], \\
Z_5 &=& 1-{\as\over4\pi}\,\CF\Big(\,4-5\,\eps\,\Big),
\eea
where the $\eps$ in $Z_5$ generates spurious finite contributions when
multiplied with the IR pole terms in the virtual- and real-gluon parts.
However, these finite terms cancel in the total results.

In the intermediate step of the derivation, we get after eliminating the
UV divergences with $Z_2$ the relation
\be\label{rhs}
Z_2\,\Big(\,1+C_{\mbox{\tiny bare}}\,\Big) =
1+C+{\as\over2\pi}\,\CF+\delta C,
\ee
where $C$ is the chromomagnetic form factor for dimensional regularization
with anticommuting $\g5$, namely Eq.~(\ref{ABC}). By multiplying the right-hand
side of Eq.~(\ref{rhs}) with $Z_5$, we finally obtain the fully renormalized
result (neglecting imaginary parts)
\bea
C_{\mbox{\tiny rep}} &=& C+\delta C-{\as\over2\pi}\CF \nn
&=& {\as\over 4\pi} \CF\Bigg[\,2\Bigg( {1+v^2\over v}
   \ln\Bigg({1+v\over 1-v}\Bigg)-2\Bigg)\Bigg(\ln\Lambda^{1\over2}
   -\Frac{1}{2}\ln(1-v^2)+\ln2+1\Bigg) \nn && \kern40pt
   -4v\ln\Bigg({1+v\over1-v}\Bigg)+F(v)+4\,\Bigg],
\eea
which agrees with the result Eqs.~(\ref{ABC}) and (\ref{C5}).
\newpage
\section*{Appendix B:\hskip4mm{
Properties of \boldmath$q\bar{q}g$} Phase-Space Integrals}
\setcounter{equation}{0}
\def\theequation{B\arabic{equation}}

The massive three-particle phase-space is as usual defined by
\be
\mbox{PS}_3 = \int\!\!
\left[\,\prod_{i=1}^3{d^N p_i\over(2\pi)^N}\,\delta^+(p_i^2-m_i^2)\,\right]\,
(2\pi)^{N+3}\delta\left(q-\sum_{i=1}^3 p_i\right),
\ee
where in this particular case of fermion pair production $m_1=m_2=m$. Further,
we put $m_3=\sqrt{q^2}\,\Lambda^{1\over2}$ and $N=4$. Regularizing by a
small gluon mass has the advantage to shift all complications to the
boundary functions of the integrals whereas dimensional regularization
produces an $\eps$ dependent integration measure which does in general not
permit to fully exploit powerful substitution techniques. It appears that
the most complicated of these phase-space integrals are only solvable in
four dimensions.

Integrating over the gluon momentum and all the internal angles except for
$\chi$ gives
\be
\mbox{PS}_3 =
{1\over(4\pi)^3}\int\!\!
\left[\,\prod_{i=1}^2{d|{\bm p}_i|\,|{\bm p}_i|^2\over\strut
\sqrt{{\bm p}_i^2+m^2}}\,\right]\;\int\limits_0^\pi\!
d\chi\,\sin\chi\ \delta\Bigl((q-p_1-p_2)^2\Bigr),
\ee
where the delta function expresses energy-momentum conservation
\bea
\delta\Bigl((q-p_1-p_2)^2\Bigr) &=&
{4\over\strut q^2\sqrt{(1-y)^2-\xi}\sqrt{(1-z)^2-\xi}}\ \times \nn[.3cm]
& & \delta\left(\,\cos\chi -
{yz+y+z+\xi-1+2\Lambda\over\strut\sqrt{(1-y)^2-\xi}\,\sqrt{(1-z)^2-\xi}}
\,\right).
\eea
Note that we do not put $\Lambda=0$ in the argument of the delta function
which directly effects the deformation of the phase-space boundaries to
regulate the IR singularities.

Finally, we obtain with the $(y,z)$-parametrization given by Eq.~(\ref{yz})
for the three-body phase-space
\be
\mbox{PS}_3 =
{q^2\over 2^7\pi^3}
\int\limits_{y_-}^{y_+}\!dy\int\limits_{z_-(y)}^{z_+(y)}\!\!dz,
\ee
where the upper and lower bounds of the nested integral follow directly from the
constraint $-1\le\cos\chi\le +1$ and Eq.~(B3)
\be
\begin{array}{l}
   \left\{
   \begin{array}{ccl}
    y_+ &=& 1-\sqrt{\xi}, \\
    y_- &=& \Lambda^{1\over 2}\,\sqrt{\xi}+\Lambda,
    \end{array}\right. \\[.6cm]
z_\pm(y) = \displaystyle {2y\over 4y+\xi}\left[\,
      1-y-\Frac{1}{2}\xi+\Lambda+{\Lambda\over y}\pm
      {1\over y}\sqrt{(1-y)^2-\xi}\,\sqrt{(y-\Lambda)^2-\Lambda\xi}\,\right].
\end{array}
\ee

In the calculation of the differential rates one requires the solutions of
the following three distinct types of phase-space integrals
\bea
\I{i} &= \uint f_i(y,z), \qquad \S{i} &= \pint f_i(y,z), \nn
\J{i} &= \jint f_i(y,z), &
\eea
where $f_i(y,z)$ are real rational functions in the quark-energy variables
$y=1-p_1\!\cdot q/q^2$ and $z=1-p_2\!\cdot q/q^2$. Apart from the numerical
verification of each individual analytical solution there exist several
consistency checks among the different integral classes. For brevity, we will
only discuss relations among $\{\I{i}\}$ and $\{\S{i}\}$, but similar
considerations apply to $\{\J{i}\}$.

It is important to recognize that although $\{\S{i}\}$ and $\{\J{i}\}$
emerge as new integrals (in addition to $\{\I{i}\}$) in the transition from the
the total cross section to the differential form of the cross section, this
procedure can not fundamentally alter the divergence structure in the soft
divergences~\cite{KLN}. Consider the following argument: The integrand
$f(y,z)$ of a divergent $\EuScript S$ is first integrated over $z$ to obtain
$F(y)$. Then, before executing the remaining $y$-integration, we add and
substract a suitable function $\tilde{F}(y)$:
\bea
\EuScript S &=& \pint f(y,z) =
{\displaystyle\int{dy\over\sqrt{(1-y)^2-\xi}}\:\:} F(y) \nn
   &=& \int dy\:\:\tilde{F}(y) +
{\displaystyle\int{dy\over\sqrt{(1-y)^2-\xi}}\:\:}
\left[\,F(y)-\sqrt{(1-y)^2-\xi}\,\tilde{F}(y)\,\right]. \label{arg1}
\eea
Now $\tilde{F}(y)$ can be choosen that way that the second integral on the
right-hand side of Eq.~(\ref{arg1}) becomes regular as $\Lambda\to0$ and all
soft divergences are contained in the first much simpler integral of type
$\EuScript I$. It becomes clear that $\EuScript S$ and $\EuScript I$ exhibit
the same divergence structure in $\Lambda$, {\it i.e.} for any $\xi\neq0$
\be
\EuScript S(\ln\Lambda) = \EuScript I(\ln\Lambda) +
\mbox{finite terms as}\ \Lambda\to 0.
\ee
On the other hand, the collinear singularities provide another useful tool
to establish a correspondence between $\EuScript S$- and
$\EuScript I$-integrals. For $\Lambda\neq0$, we can in general find a suitable
combination of ${\EuScript S}_i$ and ${\EuScript S}_k$ with $i\neq k$ so that
\be
{\EuScript S}_i(\ln\xi,1/\xi)-{\EuScript S}_i(\ln\xi,1/\xi) =
{\EuScript I}_i(\ln\xi,1/\xi) + \mbox{finite terms as}\ \xi\to0.
\ee
Furthermore, simple algebraic manipulations yield valuable identities that
save considerable labor, {\it e.g.} the following substitution was useful
in the derivation of the real-tree graph contributions
\be
\int{dy\over(1-y)^2-\xi}{y^2\over z} = \I{2}+2\J{12}-(1-\xi)\,\J{8}.
\ee

An exhaustive list with all three-body phase-space integrals relevant to the
derivation of the \oas\ differential cross section for massive fermion
pair production is given below. The integral sets $\{\I{i}\}$ and $\{\S{i}\}$
with the exception of $\S{13}$ have been published before~\cite{long}. The
integral class $\{\J{i}\}$ is entirely new. Some of its components are also
important in the angular dependence of the quark's alignment
polarization~\cite{wip}. For the $\EuScript J$-integrals we use the shorthand
$w=\sqrt{(1-\sxi)/(1+\sxi)}$. The standard book on dilogarithms and their
integral representations is Ref.~\cite{lewin}.
\def\next{\nn[1.3cm]}
\subsection*{Class \boldmath{$\EuScript I$} Integrals}
      \begin{eqnarray}
      \I{1} & = &
      \uint \nonumber \\
      & = &
      \Frac{1}{2}v\left(1+\Frac{1}{2}\xi\right)-\Frac{1}{2}\xi\left(
      1-\Frac{1}{4}\xi\right)\ln\left({1+v\over 1-v}\right)
      \next
      \I{2} & = &
      \uint {1\over y}\ =\ \uint {1\over z} \nonumber \\
      & = &
      -v+\left(1-\Frac{1}{2}\xi\right)\ln\left({1+v\over 1-v}\right)
      \next
      \I{3} & = &
      \uint {1\over y^2}\ =\ \uint {1\over z^2} \nonumber \\
      & = &
      -{4v\over\xi}\left(\ln\Lambda^{1\over 2}+\ln\xi-2\ln v-2\ln 2+1\right)+
      2\left(1-{3\over\xi}\right)\ln\left({1+v\over 1-v}\right)
      \next
      \I{4} & = &
      \uint {y\over z}\ =\ \uint {z\over y} \nonumber \\
      & = &
      -\Frac{1}{4}v\left(5-\Frac{1}{2}\xi\right)+\Frac{1}{2}\left(
      1+\Frac{1}{8}\xi^2\right)\ln\left({1+v\over 1-v}\right)
      \next
      \I{5} & = &
      \uint {1\over y z} \nonumber \\
      & = &
      \left(-2\ln\Lambda^{1\over 2}-\ln\xi+4\ln v+2\ln 2\right)
      \ln\left({1+v\over 1-v}\right) \nonumber \\
      & &
     +2\left[\Li\left({1+v\over 2}\right)-\Li\left({1-v\over 2}\right)\right]
     +3\left[\Li\left(-{2v\over 1-v}\right)-\Li\left({2v\over 1+v}\right)\right]
      \nonumber
      \end{eqnarray}
\subsection*{Class \boldmath{$\EuScript S$} Integrals}
      \begin{eqnarray}
      \S{1} & = &
      \pint \nonumber \\
      & = &
      1-\sqrt{\xi}-\Frac{1}{2}\xi\ln\left({2-\sqrt{\xi}\over\sqrt{\xi}}\right)
      \next
      \S{2} & = &
      \pint {1\over y} \nonumber \\
      & = &
      2 \ln\left({2-\sqrt{\xi}\over\sqrt{\xi}}\right)
      \next
      \S{3} & = &
      \pint {1\over y^2} \nonumber \\
      & = &
      {4\over\xi} \Biggl[ -\ln\Lambda^{1\over 2}+\Frac{1}{2}\ln\xi+
      \ln(1-\sqrt{\xi})-2\ln(2-\sqrt{\xi})+\ln 2-1 \Biggr]
      \next
      \S{4} & = &
      \pint {1\over z} \nonumber \\
      & = &
      \Li\left({1+v\over 2}\right)+\Li\left({1-v\over 2}\right)+
      2\,\Li\left(-{\sqrt{\xi}\over 2-\sqrt{\xi}}\right)+\Frac{1}{4}
      \ln^2\xi \nonumber \\
      & &
      +\ln^2\left({2-\sqrt{\xi}\over 2}\right)-\ln(1+v)\ln(1-v)
      \next
      \S{5} & = &
      \pint {1\over z^2} \nonumber \\
      & = &
      {4\over\xi} \left[ -\ln\Lambda^{1\over 2}-\Frac{1}{2}\ln\xi+
      \ln(1-\sqrt{\xi})-{1+v^2\over 2v}
      \ln\left({1+v\over 1-v}\right)+\ln 2 \right]
      \next
      \S{6} & = &
      \pint {y\over z^2} \nonumber \\
      & = &
      {4\over\xi} \left(1-\sqrt{\xi}\right) % \nonumber
      \next
%
%     \end{eqnarray}
%     \eject\newpage
%     \begin{eqnarray}
%
      \S{7} & = &
      \pint {y^2\over z^2} \nonumber \\
      & = &
      {2\over\xi} \left(1-\sqrt{\xi}\right)^2
      \next
      \S{8} & = &
      \pint z \nonumber \\
      & = &
      {1\over 32} \left[ 12-(2+\xi)^2-{2+\sqrt{\xi}\over 2-\sqrt{\xi}}\,\xi^2
      +2(8-\xi)\xi\,\ln{\sqrt{\xi}\over 2-\sqrt{\xi}} \right]
      \next
      \S{9} & = &
      \pint {z\over y} \nonumber \\
      & = &
      -\Frac{1}{2}\ln\xi+\ln\left(2-\sqrt{\xi}\right)+
      {2\over 2-\sqrt{\xi}}-2
      \next
      \S{10} & = &
      \pint {y\over z} \nonumber \\
      & = &
      \Li\left({1+v\over 2}\right)+\Li\left({1-v\over 2}\right)-
      2\,\Li\left({\sqrt{\xi}\over 2}\right)+\Frac{1}{4}
      \ln^2\bigl(\Frac{1}{4}\xi\bigr) \nonumber \\
      & &
      +\left(2-\Frac{1}{2}\xi\right)
      \ln\left({2-\sqrt{\xi}\over\sqrt{\xi}}\right)-\sqrt{\xi}+
      2v\ln\left({1-v\over 1+v}\right) \nonumber \\
      & &
      -\ln\left({1+v\over 2}\right)\ln\left({1-v\over 2}\right)+1
      \next
      \S{11} & = &
      \pint {y^2\over z} \nonumber \\
      & = &
      \left(1+\Frac{1}{2}\xi\right) \left[ \Li\left({1+v\over 2}\right)+
      \Li\left({1-v\over 2}\right)-2\,\Li\left(\Frac{1}{2}\sqrt{\xi}\right)+
      \Frac{1}{4}\ln^2\left(\Frac{1}{4}\xi\right) \right. \nonumber \\
      & &
      \left. -\ln\left({1+v\over 2}\right)\ln\left({1-v\over 2}\right) \right]+
      3v \ln\left({1-v\over 1+v}\right)+
      \Frac{1}{8}(18+\xi)-\Frac{1}{8}(20-\xi)\sqrt{\xi} \nonumber \\
      & &
      +\left(3-\xi+\Frac{1}{16}
      \xi^2\right) \ln\left({2-\sqrt{\xi}\over\sqrt{\xi}}\right)
%     \nonumber
      \next
%
%     \end{eqnarray}
%     \eject\newpage
%     \begin{eqnarray}
%
      \S{12} & = &
      \pint {1\over y z} \nonumber \\
%     & & \nonumber \\
%     & & \nonumber \\
      & = &
      {1\over v}\ln\left({1-v\over 1+v}\right) \left[ 2\ln\Lambda^{1\over 2} +
      \Frac{1}{2}\ln\xi+4\ln(2-\sqrt{\xi})-4\ln v-4\ln 2-
      2\ln\left({1-v\over 1+v}\right) \right] \nonumber \\
%     & & \nonumber \\
      & &
      + {1\over v}\ln^2\left({(1-v)^2\over\sqrt{\xi}(2-\sqrt{\xi})}\right)+
      {2\over v} \ln\left({\sqrt{\xi}(2-\sqrt{\xi})\over 2}\right)
      \ln\left({2\sqrt{\xi}(1-\sqrt{\xi})\over (1-\sqrt{\xi}-v)^2}\right)
      \nonumber \\
      & & \nonumber \\
      & &
      +{2\over v} \left[ \Li\left({\sqrt{\xi}(2-\sqrt{\xi})\over (1+v)^2}
      \right)-\Li\left[\left({1-v\over 1+v}\right)^2\right]+
      \Li\left({(1-v)^2\over\sqrt{\xi}(2-\sqrt{\xi})}\right) \right]
      \nonumber \\
      & & \nonumber \\
      & &
      +{1\over v} \left[ \Li\left({1+v\over 2}\right)-
      \Li\left({1-v\over 2}\right)+\Li\left(-{2v\over 1-v}\right)-
      \Li\left({2v\over 1+v}\right)-{\pi^2\over 3} \right]
      \nonumber \\
      \nonumber \\
      \S{13}& = &
      \pint y \nonumber \\
      & = &
      {1\over 16} \left[ -\xi^2\ln\xi+2\,\xi^2\ln(2-\sqrt{\xi})+
      4(2-\sqrt{\xi})^2-4\left(2-\xi^{3\over2}\right)\right]
      \nonumber
      \end{eqnarray}
\subsection*{Class \boldmath{$\EuScript J$} Integrals}
      \begin{eqnarray}
      \J{1} & = &
      \jint \nonumber \\
      & = &
      2\,{1-\xi\over 4-\xi}\ln\left({1+v\over 1-v}\right)
      \next
      \J{2} & = &
      \jint {1\over y} \nonumber \\
      & = &
      {6\over 4-\xi}\ln\left({1+v\over 1-v}\right)
      \next
      \J{3} & = &
      \jint {1\over y^2} \nonumber \\
      & = &
      {4\over\xi\,v}\Biggl[\,-\ln\Lambda^{1\over 2}
      -\ln\xi+2\ln v+2\ln2-1\,\Biggr]
      -{24\over\xi(4-\xi)}\ln\left({1+v\over 1-v}\right)
      \next
      \J{4} & = &
      \jint y \nonumber \\
      & = &
      -\left(1+\Frac{1}{2}\xi-{6\over 4-\xi}\right)
      \ln\left({1+v\over 1-v}\right)-v
      \next
      \J{5} & = &
      \jint y^2 \nonumber \\
      & = &
      \left(2+\Frac{1}{2}\xi+\Frac{1}{8}\xi^2-{6\over 4-\xi}\right)
      \ln\left({1+v\over 1-v}\right)-\Frac{1}{4}(6-\xi)v
      \next
      \J{6} & = &
      \jint y z \nonumber \\
      & = &
      \Frac{1}{2}\xi\left(-2+\Frac{1}{8}\xi+{9\over 4-\xi}-
      {12\over(4-\xi)^2}\right)\ln\left({1+v\over 1-v}\right)+
      \left(\Frac{3}{4}+\Frac{1}{8}\xi-{2\over 4-\xi}\right)v
      \next
      \J{7} & = &
      \jint z \nonumber \\
      & = &
      -{3\xi\over 4-\xi}\left(1-{2\over 4-\xi}\right)
      \ln\left({1+v\over 1-v}\right)+
      {2v\over 4-\xi}
      \next
      \J{8} & = &
      \jint {\:1\over z} \nonumber \\
      & = &
      \frac1{\sxi}\Bigg[\,\Li(w)-\Li(-w)
      +\Li\left({2+\sxi\over 2-\sxi}\;w\right)
      -\Li\left(-{2+\sxi\over 2-\sxi}\;w\right)\,\Biggr]
      \next
      \J{9} & = &
      \jint {1\over z^2} \nonumber \\
      & = &
      {4\over\xi\,v}\Biggl[\,-\ln\Lambda^{1\over 2}
      -\ln\xi+2\ln v+2\ln2
      -\frac{2-\xi}{2v}\ln\left(\frac{1+v}{1-v}\right)\,\Biggr]
      \next
      \J{10} & = &
      \jint {1\over yz} \nonumber \\
      & = &
      {1\over 2(1-\xi)}\ln\left(\frac{1+v}{1-v}\right)
      \Biggl[-4\ln\Lambda^{\frac12}-\Frac{5}{2}\ln\xi+5\ln(1+\sxi)+4\ln(1-\sxi)
      \nonumber \\ &&
      \hphantom{-\ln(2+\sxi)+6\ln2\Biggr]}-\ln(2+\sxi)+6\ln2\Biggr]
      \nonumber \\ &&
      +{2\over 1-\xi}\Biggl[\,\Li\left(\frac{1+v}2\right)
      -\Li\left(\frac{1-v}2\right)\,\Biggr]
      +{3\over 1-\xi}\Biggl[\,\Li\left(-{2v\over 1-v}\right)
      -\Li\left(\frac{2v}{1+v}\right)\,\Biggr]
      \nonumber \\ &&
      +{1\over\sxi(1-\sxi)}\Bigg[\,\Li(w)-\Li(-w)
      +\Li\left({2+\sxi\over 2-\sxi}\;w\right)
      -\Li\left(-{2+\sxi\over 2-\sxi}\;w\right)\,\Biggr]
      \nonumber \\ &&
      -{1\over 1-\xi}\Bigg[
      \Li\left({1+w\over 2}\right)-\Li\left({1-w\over 2}\right)
      +\Li\left((2+\sxi){1+w\over 4}\right)
      \nonumber \\ &&
      -\Li\left((2+\sxi){1-w\over 4}\right)
      +\Li\left({2\sxi\over(2+\sxi)(1+w)}\right)
      -\Li\left({2\sxi\over(2+\sxi)(1-w)}\right)
      \Bigg]
      \next
      \J{11} & = & \jint {z\over y} \nonumber \\
      & = &
      \left(-1+{12\over 4-\xi}-{24\over(4-\xi)^2}\right)
      \ln\left({1+v\over 1-v}\right)
      -\frac{2v}{4-\xi}
      \next
      \J{12} & = &
      \jint {y\over z} \nonumber \\
      & = &
      \Frac{1}{2}\ln\left(\frac{1+v}{1-v}\right)
      \Biggl[\,\Frac{1}{2}\ln\xi+\ln(2+\sxi)-\ln(1+\sxi)-2\ln2\,\Biggr]
      \nonumber \\ &&
      +{1-\sxi\over\sxi}\Bigg[\,\Li(w)-\Li(-w)
      +\Li\left({2+\sxi\over 2-\sxi}\;w\right)
      -\Li\left(-{2+\sxi\over 2-\sxi}\;w\right)\,\Biggr]
      \nonumber \\ &&
      +\Li\left({1+w\over 2}\right)-\Li\left({1-w\over 2}\right)
      +\Li\left((2+\sxi){1+w\over 4}\right)
      -\Li\left((2+\sxi){1-w\over 4}\right)
      \nonumber \\ &&
      +\Li\left({2\sxi\over(2+\sxi)(1+w)}\right)
      -\Li\left({2\sxi\over(2+\sxi)(1-w)}\right)
      \next
      \J{13} & = &
      \jint {y\over z^2} \nonumber \\
      & = &
      \frac2\xi\ln\left({1+v\over 1-v}\right)
      \next
      \J{14} & = &
      \jint {y^2\over z^2} \nonumber \\
      & = &
      \frac2\xi\left[\,\ln\left({1+v\over 1-v}\right)-2v\,\right]
      \next
      \J{15} & = &
      \jint {y^3\over z} \nonumber \\
      & = &
      \Frac{1}{2}(3+\xi)\ln\left({1+v\over 1-v}\right)
      \Biggl[\,\Frac{1}{2}\ln\xi+\ln(2+\sxi)
      -\ln(1+\sxi)-2\ln2\,\Biggr]
      \nonumber \\ &&
      +{(1-\sxi)^3\over\sxi}\Bigg[\,\Li(w)-\Li(-w)
      +\Li\left(\frac{2+\sxi}{2-\sxi}\;w\right)
      -\Li\left(-\frac{2+\sxi}{2-\sxi}\;w\right)\,\Biggr]
      \nonumber \\ &&
      +(3+\xi)\Bigg[\,\Li\left({1+w\over 2}\right)
      -\Li\left({1-w\over 2}\right)
      +\Li\left((2+\sxi){1+w\over 4}\right)
      \nonumber \\ &&
      -\Li\left((2+\sxi){1-w\over 4}\right)
      +\Li\left(\frac{2\sxi}{(2+\sxi)(1+w)}\right)
      -\Li\left(\frac{2\sxi}{(2+\sxi)(1-w)}\right)\,\Bigg]
      \nonumber \\ &&
      +{40-16\xi+\xi^2\over16}\ln\left(\frac{1+v}{1-v}\right)
      -{26-\xi\over 8}\,v
      \next
      \J{16} & = &
      \jint {y^3\over z^2} \nonumber \\
      & = &
      {1\over\xi}\Biggl[\,(2+\xi)\ln\left({1-v\over 1+v}\right)+6v\,\Biggr]
      \nonumber
      \end{eqnarray}
\newpage
\newpage
\centerline{\bf\Large Figure Captions }
\vspace{1cm}
\newcounter{fig}
\begin{list}{\bf\rm Fig.\ \arabic{fig}: }{\usecounter{fig}
\labelwidth1.6cm \leftmargin2cm \labelsep0.4cm \itemsep0ex plus0.2ex }

\item Feynman diagrams contributing to
      $d\sigma/d\!\ct\;(e^+ e^-\to q\bar{q})$
      up to order \oas:
      (a) Born term, (b) virtual corrections,
      and (c) gluon bremsstrahlung graphs.
      Diagrams (b) and (c) give the definition
      of the particle momenta.
\item Two-particle kinematics for the
      process $e^+(p_+)\,e^-(p_-)\to q(p_1)\,\bar{q}(p_2)$
      in the cms coordinate frame.
      The $e^+ e^-$ beam line coincides with the $z$-axis.
      The scattering angle between the electron momentum {\boldmath $p_-$}
      and the quark momentum {\boldmath $p_1$} is $\theta$, and angle
      $\varphi$ defines the orientation around the beam axis.
\item Three-particle kinematics for the
      process $e^+(p_+)\,e^-(p_-)\to q(p_1)\,\bar{q}(p_2)\,g(p_3)$
      in the cms frame.
      The momentum definitions agree with those of Fig.~2
      except for the additional gluon momentum $\boldmath p_3$. The
      vectors ({\boldmath $p_1$}, {\boldmath $p_2$}, {\boldmath $p_3$})
      span the $q\bar{q}g$ production plane. The angle
      $\chi\equiv\angle(\mbox{\boldmath $p_1$}, \mbox{\boldmath $p_1$})$
      and the quark momentum {\boldmath $p_1$} uniquely specify all the
      remaining momenta within the plane.
\item Differential cross section for bottom quark production: (a) \oas\ result
      as a function of the cms energy and $\ct$, and (b) the \oas\ corrections
      compared to the Born approximation. Shown are the values for the ratio
      $\Big[\,d\sigma(\mbox{\oas})/d\!\ct\,\Big]/
      \Big[\,d\sigma(\mbox{Born})/d\!\ct\,\Big]-1$
      in percent. The bottom quark mass is $m_b(m_b)=4.3\,$GeV.
\item \oas\ Differential cross section for top quark production with a top mass
      of $m_t(m_t)=174\,$GeV: (a) surface plot to show functional dependence
      on the cms energy and scattering angle, and (b) the \oas\ corrections
      compared to the Born approximation. At $375\,$GeV the corrections
      amount to $30\,\%$.

\end{list}

\begin{thebibliography}{99}
\bibitem{LEP}    The LEP Collaborations and LEP Electroweak Working Group,
                 CERN Report No.\ CERN-PPE/94-100, 1994 (unpublished), and
                 references therein. 
\bibitem{SLC}    The SLD and SLC Collaborations, in {\it Proceeedings of the
                 2nd International Workshop on Physics and Experiments with
                 Linear $e^+ e^-$ Colliders}, Waikoloa, Hawaii, 1993, edited
                 by F.A.~Harris {\it et al.} (World Scientific, Singapore,
                 1993), and references therein.
\bibitem{grun}   G.~Grunberg, Y.J.~Ng, and S.-H.H. Tye, Phys.\ Rev.\ D
                 {\bf 21}, 62 (1980).
\bibitem{JLZ}    J.~Jer\'sak, E.~Laermann and P.M.~Zerwas, Phys.\ Lett.\ B
                 {\bf 98}, 363 (1981); Phys.\ Rev.\ D {\bf 25} 1218 (1982).
\bibitem{ABL}    A.B.~Arbuzov, D.Y.~Bardin, and A.~Leike, Mod.\ Phys.\
                 Lett.\ A {\bf 7}, 2029 (1992); erratum-{\it ibid.\/}
                 A {\bf 9}, 1515 (1994).
\bibitem{thesis} M.M.~Tung, Ph.D.\ thesis, University of Mainz, 1993.
\bibitem{long}   J.G.~K\"orner, A.~Pilaftsis and M.M.~Tung, Z.\ Phys.\ C
                 {\bf 63}, 575 (1994), e-preprint hep-ph/9311332.
\bibitem{approx} M.M.~Tung, Phys.\ Rev.\ D {\bf 52}, 1353 (1995), e-preprint
                 hep-ph/9403322.
\bibitem{beam}   S.~Groote, J.G.~K\"orner and M.M.~Tung, {\it Longitudinal
                 Contribution to the Alignment Polarization of Quarks Produced
                 in $e^+ e^-$-Annihilation: An \oas\ Effect}, preprint no.\
                 MZ-TH/95-09 and FTUV/95-13, July 1995, to be published in
                 Z.\ Phys.\ C, e-preprint hep-ph/9507222.
\bibitem{muta}   T.~Muta, {\it Foundations of Quantum Chromodynamics}
                 (World Scientific, Singapore, 1987), p.\ 335.                  
\bibitem{CFH}    M.~Chanowitz, M.~Furman, and I.~Hinchcliffe, Nucl.\ Phys.\
                 B {\bf 159}, 225 (1979).
\bibitem{red}    J.G.~K\"orner and M.M.~Tung, Z.\ Phys.\ C {\bf 64}, 255 (1994).
\bibitem{thooft} G.~'t~Hooft and M.~Veltman, Nucl.\ Phys.\ B {\bf 44},
                 189 (1972).
\bibitem{AD}     D.~Akyeampong and R.~Delbourgo, Nuevo Cimento {\bf 17 A},
                 578 (1973).
\bibitem{larin}  S.A.~Larin, Phys.\ Lett.\ B {\bf 303}, 113 (1993),
                 e-preprint hep-ph/9302240.
\bibitem{LV}     S.A.~Larin and J.A.M.~Vermaseren, Phys.\ Lett.\ B {\bf 259},
                 345 (1991).
\bibitem{KLN}    T.~Kinoshita, J.\ Math.\ Phys. {\bf 3}, 650 (1962);
                 T.D.~Lee and M.~Nauenberg, Phys.~Rev.\ {\bf 133}, 1549 (1964).
\bibitem{run}    W.~Bernreuther, in {\it Proceedings of the Workshop on QCD at
                 LEP 1994\/}, Aachen, Germany, 1994;
                 G.~Rodrigo and A.~Santamaria, Phys.\ Lett.\ B {\bf 313},
                 441 (1993), e-preprint hep-ph/9305305.
\bibitem{wip}    S.~Groote, J.G.~K\"orner and M.M.~Tung, {\it Polar Dependence
                 of the Alignment Polarization of Quarks Produced in
                 $e^+ e^-$-Annihilation\/}, preprint no.\
                 MZ-TH/95-19 and FTUV/95-52, Januar 1996.
\bibitem{PV}     G.~Passarino and M.~Veltman, Nucl.\ Phys.\ B {\bf 160},
                 151 (1979).
\bibitem{lewin}  L.~Lewin, {\it Polylogarithms and Associated Functions}
                 (North Holland, New York, 1981); for a recent compilation
                 of dilogarithm identities see also: A.N.~Kirillov, Supp.\
                 Prog.\ Theo.\ Phys.\ {\bf 118}, 61 (1995).
\bibitem{bodo}   A.~Djouadi, B.~Lampe, and P.M.~Zerwas, Z.\ Phys.\ C
                 {\bf 67}, 123 (1995), e-preprint hep-ph/9411386.

\end{thebibliography}
\end{document}